\documentclass[paper]{JHEP}
\usepackage{amssymb}
\usepackage{epsfig}
\usepackage{eucal}
\usepackage{amsmath}
\usepackage{enumerate}
\renewcommand{\nc}{\newcommand}
\nc{\cp}{\boldmath{$CP$}}
\nc{\tb}{\bar{t}}
\nc{\pt}{p_t}
\nc{\ptb}{p_{\bar{t}}}
\nc{\st}{s_t}
\nc{\stb}{s_{\bar{t}}}
\nc{\pA}{p_A}
\nc{\pB}{p_B}
\nc{\pa}{p_a}
\nc{\pb}{p_b}
\nc{\g}{\mbox{\sl g}}
\nc{\shat}{\hat{s}}
\nc{\that}{\hat{t}}
\nc{\uhat}{\hat{u}}
\nc{\htheta}{\hat{\theta}}
\nc{\non}{\nonumber}
\nc{\ktil}{\tilde{\kappa}}
\nc{\mt}{m_t}
\nc{\costhhat}{\cos\hat{\theta}}
\nc{\betat}{\beta_t}
\nc{\gammat}{\gamma_t}
\nc{\LT}{\left}
\nc{\RT}{\right}

\preprint{KOBE-FHD-01-03\\
hep-ph/0105117}

\title{\boldmath{$CP$} Violation via Top Quark Anomalous Interaction
       at the Fermilab Tevatron}
\author{\sc Kazumasa OHKUMA \\ 
Graduate School of Science and Technology,
Kobe University \\  {\sl Nada, Kobe 657-8501, JAPAN.}\\
E-mail: \email{ohkuma@radix.h.kobe-u.ac.jp}}

\abstract{
Expecting the forthcoming experiment at 
the upgraded Fermilab Tevatron,
we calculated  $C\!P$-violating polarization asymmetry of 
$t\bar{t}$, 
${\cal A}_{CP}\equiv [\: \sigma(p\bar{p}\rightarrow t(-) \bar{t}(-) X)
-\sigma(p\bar{p}\rightarrow t(+) \bar{t}(+) X)\:]~/~
\sigma(p\bar{p}\to t\bar{t}X)$,
due to possible anomalous chromomagnetic ($\kappa$) 
and chromoelectric ($\ktil$) couplings of the gluon to 
the top quark.
Since $\kappa$ and $\ktil$ are sensitive to a contribution
from new physics beyond standard model,
this observable is useful to search for a signal of new physics.
It was seen that the magnitude of 
${\cal A}_{CP}$ depends only  on Im($\ktil$) and
Im($\kappa^* \ktil$).
Futhermore, we found that 
when $|$Im($\kappa^* \ktil$)$|>0.5$,
one can possibly detect the $C\!P$-violation
effect as a signal of new physics even if the magnitude of 
Im($\ktil$) is zero.
}

\begin{document}

\section{INTRODUCTION}
\label{sec:intro}
The Standard model (SM) which is composed of electroweak theory and quantum 
chromodynamics is  extremely successful in particle physics phenomenology. 
The predictions by the SM have been in agreement with all experimental data 
up to the scale of $O(M_{W/Z})$.
Recently Brookhaven E821 group reported that 
observed anomalous magnetic moment of positive muon 
was in disagreement 
with the SM prediction at 2.6 standard deviation~\cite{g-2},
which seems to suggest the existence of new physics beyond SM.
However, this experiment was  already closed,
though the statistical error might become smaller 
in undergoing  data analysis~\footnote{New mesurements are now underway with
negative muon which will provide a sensitive test of $C\!P\!T$-violation.}.  
Therefore, it is very important and challenging to search for a signal of 
new physics at forthcoming high energy collider experiments.

On the other hand, the physics of the top quark which is discovered 
as a very heavy particle, $\mt~\simeq$ 180 GeV,
at the Fermilab Tevatron
with center-of-mass energy $\sqrt{s}$\ =\ 1.8\ TeV 
in 1994~\cite{cdf},
is very interesting.
Since the mass of the top quark is much larger 
than the masses of  other
quarks and leptons (and even those of the electroweak gauge bosons),
studies on the role of this particle in Nature is expected to lead us to 
the physics  beyond the SM.
For example,
as it is well-known,
the $C\!P$-violation in the top quark pair production  
is estimated to be extremely  small in the SM~\cite{smtopcp},
based on the Cabibbo-Kobayashi-Maskawa (CKM) mechanism~\cite{cabibbo,kobayashi}.
Thus, it is very interesting to search for 
other possible origin of $C\!P$-violation in the  top quark sector
originated from new physics.
Furthermore, it is remarkable that
due to its huge mass, 
the top quark decays before it
hadronizes~\cite{bigi}.
Then,  we can easily get 
information about physics of the produced top quark
from the decay distribution of secondary leptons 
and hadrons~\cite{tdecay,cedm1}.
These properties  are very advantageous for searching the  $C\!P$-violation 
in top quark sector as a signal of new physics. 

Based on the above consideration,
in this paper we investigate the $C\!P$-violation effect 
originated from new physics
in the top quark pair production process 
at the upgraded Fermilab Tevatron.
Here, we assume the existence of chromomagnetic and chromoelectric type
couplings which could  be sizable if new physics is in existence
as a non-standard effect.

This paper is organized as follows.
In Chapter~\ref{sec:cpvt}, 
we introduce the $C\!P$-violating observable
for the top quark pair production and the effective Lagrangian which
include anomalous chromomagnetic and chromoelectric couplings of gluon
to top quark.
In Chapter~\ref{sec:analysis}, we calculate the effect of
the $C\!P$-violation for $p\bar{p} \rightarrow t \bar{t}X $
processes and discuss the results.
Chapter~\ref{sec:summary} is devoted to the summary of 
this work and a future outlook.

\section{{\mbox{\boldmath{$C\!P$}}}-VIOLATION 
IN TOP QUARK PAIR PRODUCTION}
\label{sec:cpo}
In this section, to study the contribution 
of the top quark anomalous chromomagnetic 
and chromoelectric dipole type couplings 
to the polarization asymmetry for the process;
\begin{equation}
p \bar{p}\rightarrow t \bar{t} X,
\label{eq:ppttbx}
\end{equation}
which will be observed at the upgraded Fermilab Tevatron with
$\sqrt{s} = 2.0$ TeV, we introduce an  observable for $C\!P$-violation
and an effective Lagrangian which includes
the top quark anomalous chromomagnetic 
and chromoelectric dipole type couplings.

Since the quark--anti-quark annihilation 
($q \bar{q} \to t \bar{t}$) is the major source of the top quark pair
production  at the Tevatron as shown in Fig.~\ref{fig:ppttb},
we neglect subprocesses of the gluon fusion (gg$ \to t \bar{t}$)
in this work~\cite{topevent}.

%%%%%%%%%%%%%%%%%%%%%%%%%%%%%%%%%%%%%%%%%%%%%%%%%%%%%%%%%%%%%%%%%%%%%
%                              _      _
%     Diagram of the process p p --> ttX
%
%%%%%%%%%%%%%%%%%%%%%%%%%%%%%%%%%%%%%%%%%%%%%%%%%%%%%%%%%%%%%%%%%%%%%
\FIGURE{\epsfig{file=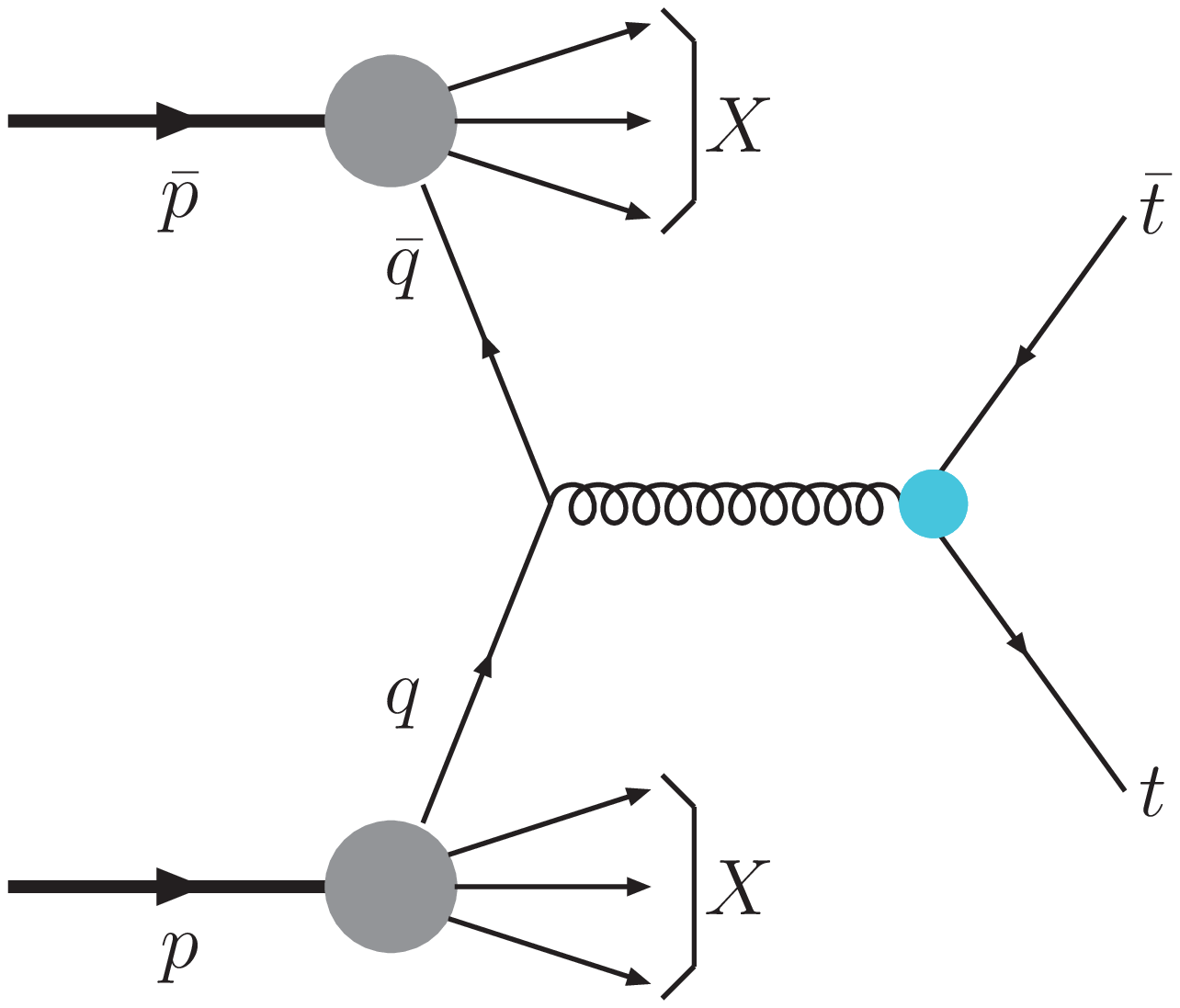,width=9cm}
\caption
{Fyenman diagram for top quark pair production at Tevatron}
\label{fig:ppttb}}

\label{sec:cpvt}
\subsection{\bf {\mbox{\boldmath{$C\!P$}}}-violating Observable}
\label{subsec:cpob}
Since top quark pairs are mainly produced through the gluon
interaction at Tevatron (Fig.~\ref{fig:ppttb}),
the helicities of $t\tb$ would be ($+ -$) or  ($- +$) 
due to the helicity conservation 
which is realized if the top quark mass is  much smaller
than $\sqrt{s}$.
However, since top quark mass is about 180 GeV, we can also expect to
have 
($+ +$) and  ($- -$) combinations as a consequence of the 
breaking of the helicity conservation.
We can use these combinations to study $C\!P$ properties of the 
 $t\tb$
state;
$|h_{t} h_{\bar{t}}\rangle$.

$|- +\rangle$ and $|+ -\rangle$ are $C\!P$ self-conjugate while
$|- -\rangle$ and $|+ +\rangle$ transform  each other under
$C\!P$ operation as
\begin{equation}
\hat{C}\hat{P}|\mp \mp\rangle 
= \hat{C} |\pm \pm\rangle
= |\pm \pm\rangle.
\end{equation}
Therefore, the difference between the events  
$N(- - )$ and $N(+ +)$ 
could be a useful measurement of $C\!P$ violation~\cite{SP};
\begin{align}
{\cal A}_{cp}&=\frac{N(- -) - N(+ +)}{N(all)} \non\\
      &=\frac{\sigma_{- -} - \sigma_{+ +}}{\sigma_{\rm total}}
      \equiv\frac{\Delta \sigma}{\sigma_{\rm total}}
      \label{eq:crossasym},\\
      & N(all)\equiv N(+ +) + N (+ -) +  N(- +) + N (- -)\non ,
\end{align}
where $\sigma_{+ +}$ and  $\sigma_{- -}$ is 
the cross section of the $t\tb$ production
with the  helicities ($+ +$) and ($- -$).

Though the produced top quark can not be directly  detected,
we can easily reconstruct the top quark signal
through the produced top quark decay distribution 
of the secondary leptons~\cite{tdecay,cedm1}.
Notice that  there are some remarkable properties 
about the top quark decays~\cite{SP};
\begin{enumerate}[(1)]
\item\label{mosc:liftime}
Since top quark is very heavy ($\mt \sim 180$ GeV) and  
its life is much shorter than $10^{-23}$ s
being smaller than  the hadronization time,
the top would decay without the hadronic effect~\cite{bigi}.   
\item\label{mosc:handedness}
Since $\mt > m_{\rm W}$, the dominant decay process of the top quark should
be  $t~\rightarrow W^+~b$. 
Because of large top mass, the  $W$ will be predominantly longitudinal
while the $b$ is always left-handed in the SM, if $m_b/\sqrt{s}<<1$
\footnote{At the collider energy which we focus in this work,
it is a good approximation to ignore the $b$ quark mass.
Therefore, charged currents for $b$ quarks become pure $V-A$,
where $V$ and $A$ mean vector and axial currents, respectively,
and hence $b$ quarks are treated as left-handed particles. 
}.
\item\label{mosc:secondlylepton}
Because of (\ref{mosc:handedness}),
a $t(-)$ will decay to an energetic $b(-)$
\footnote{For example, $t(-)$ and b(+) denote the left-handed top quark 
and right-handed bottom quark, respectively.}, 
which must go forward to carry the quark spin, 
and to a less energetic $W^+$;
for $t(+)$, the relative energies of $b$ and $W$ are roughly reversed. 
\item\label{mosc:tracking}
Therefore,  we can effectively get  information about the polarization
of the top quark by observing the energy distribution of the $W$ bosons
or their decay leptons.
\end{enumerate}

In addition, as described before
the  $C\!P$-violation in top quark pair production is 
estimated to be extremely small in the CKM mechanism~\cite{smtopcp}.
Thus, we have a good opportunity for  investigating the non-standard
origin of $C\!P$-violation in top quark sector.

In the following sections, we calculate the $C\!P$-violating 
observable for $p\bar{p}\to t \bar{t} X$ at Tevatron by using effective
Lagrangian being introduced in the following subsection.
\subsection{\bf Effective Lagrangian}
%\label{subsec:efflag}
In order to estimate the effect of $C\!P$-violation 
in the top quark pair production at Tevatron,
we take the following effective Lagrangian 
for top-quark--gluon interaction:
\begin{align}
L_{t\bar{t}{\rm g}}=g_s   T^a \bar{v}_{\bar{t}}
\left[
-\gamma^{\mu}G_{\mu}^a
-\frac{\kappa}{4\mt}\sigma^{\mu \nu} G^a_{\mu \nu}
-\frac{i\ktil}{4\mt}\sigma^{\mu \nu} \gamma_5 G^a_{\mu \nu}
\right]
u_t,
\label{eq:efflag}
\end{align}
where $g_s$,  $T^a$ and $\mt$ are the  strong coupling constant,
SU(3) color matrices and  top quark mass, respectively.
$G_{\mu\nu}^a$ means the gluon field strength and
$\sigma^{\mu\nu} \equiv i/2[\gamma^{\mu},\gamma^{\nu}]$.
The $\kappa$ and $\ktil$  are 
the chromomagnetic  and chromoelectric coupling, respectively.
Though many works have been done so far on
the anomalous chromomagnetic and chromoelectric dipole 
couplings~\cite{smtopcp,cedm1,cedm3,cedm4,cedm5},
those authors have treated them as real parameters
because they have focused only  on
the effective Lagrangian whose dimension is smaller than or equal to five.
However, since in general the anomalous chromomagnetic and chromoelectric
dipole moments can be  originated from some loop corrections,  
they can also have imaginary parts.
Therefore, we will treat them as complex numbers in our calculations.
From Eq.~\eqref{eq:efflag}, we can derive the effective couplings
of $t\bar{t}{\rm g}$ interaction:
\begin{align}
&\Gamma_{t\bar{t}{\rm g}}=-ig_sT^a \bar{v}(\ptb,\stb) 
      \left[
          \gamma^{\mu} +\frac{i \sigma^{\mu \nu}}{2 \mt}q_{\nu}
             \left(
                    \kappa + i \ktil \gamma^5
             \right)  
      \right] u(\pt, \st),
\label{eq:copttg}
\end{align}
where $p_t(\ptb)$, $\st(\stb)$  and $q_{\nu}$ denote four-momenta
of the top (anti-top) quark,
spin vector of the top (anti-top) quark and incoming gluon
momentum, respectively.
Though a ${\rm gg}t\bar{t}$ four-point interaction is also 
induced as a result of
gauge invariance, we can neglect  such an interaction
because top quark pair is dominantly produced through quark--anti-quark
annihilation processes at the upgraded Tevatron energy.
Furthermore, since we assume that there is no new physics except for the
interaction related to the top quark, we use the standard form 
of  quark--gluon couplings for ordinary quarks:
\begin{equation}
\Gamma_{q\bar{q}g}=-ig_sT^a \bar{v}(\pb)\gamma^{\mu}u(\pa),
\label{smcoup}
\end{equation}
where $\pa$($\pb$) is four-momentum of quarks  from
the initial proton (anti-proton).

It is remarkable that our approach does not depend on  the specific models
because we treat the anomalous chromomagnetic and chromoelectric couplings
as free parameters.
As is well known,  in the SM these  couplings
are estimated to be too small to be  detected at the Tevatron.
Therefore, if we detect the signal of these couplings,
it can be a good  evidence of new physics beyond the SM.
\section{CALCULATION AND DISCUSSION}
\label{sec:analysis}
First, let us  focus on the $C\!P$-violating observable 
$\Delta \sigma $, i.e.,  the numerator of Eq.~\eqref{eq:crossasym} defined by
\begin{align}
\Delta \sigma
&\equiv [\sigma(p~\bar{p}~\rightarrow ~t(-)~\bar{t}(-)~X)
-\sigma(p~\bar{p}~\rightarrow ~t(+)~\bar{t}(+)~X)],
\label{eq:deltan}
\end{align}
where $\sigma(p\bar{p}\rightarrow \dots)$ denotes
the cross section of top pair production for each helicity state
at the Tevatron.
Furthermore, 
$\Delta \sigma$ is given by  
\begin{align}
\Delta \sigma
&=\sum_{q} \int^1_0 dx_a \int^1_0 dx_b \int^{1}_{-1} 
d\cos \hat{\theta}
f_{q/p}(x_a) f_{\bar{q}/\bar{p}}(x_b)\non \\
&\times \Theta(x_a x_b s-4\mt^2)
\frac{d \Delta \hat{\sigma}}
{d \cos \hat{\theta}}J,
\end{align}
where the sum runs over quark flavors; 
$q = u, d, c, s, b$~\footnote{
We assume that  top quarks do not exist in the proton as partons.}.
$f_{q/p}(x_a), f_{\bar{q}/\bar{p}}(x_b)$ and $\Theta$  are 
parton distribution  functions and usual step function, respectively.
By using Eqs.~(\ref{eq:copttg}-\ref{smcoup}), the subprocess cross section, 
${d \Delta \hat{\sigma}}/{d \cos \hat{\theta}}$,
is calculated as
\begin{align}
\frac{d\Delta\hat{\sigma}}{d\costhhat}=
&\frac{\pi \alpha_s^2 \betat }{9\gammat \mt \shat^2 \sqrt{\shat}}
\left[{\rm Im{\ktil}}
                   \left\{
                          (2-\betat^2)\shat \cos^2\hat{\theta}
                              -4\mt^2
                           \right\} \right.\non \\ 
& \left.    - {\rm Im}(\kappa^*\ktil)\shat(1-2\cos^2\hat{\theta})
\right],
\label{eq:dsigma}
\end{align}
where
$\hat{s}\equiv x_a x_b s$,
$\betat\equiv\sqrt{1-{4 \mt^2}/{\shat}}, 
\gammat\equiv \sqrt{1- \betat^2}$.
$\hat{\theta}$ denote the emission angle of the top quark
in the parton center-of-mass system which is defined in Appendix~\ref{sec:cms}.
In addition, $J$ means the Jacobian which transforms
the variable $\hat{t}$ to $\cos \hat{\theta}$, given as
$J=\LT|
            {\hat{s}\betat}/{2}
         \RT|.
$
Notice that possibilities of the $C\!P$-violation roughly
depend only on  Im($\kappa^* \ktil$) and Im($\ktil$) in  
the $d \Delta \hat{\sigma}/d \cos \hat{\theta}$.
Therefore, it is possible to measure $C\!P$-violating 
observable $\Delta \sigma$ in some combinations of 
the magnitude of Im($\kappa^* \ktil$) 
and Im($\ktil$).

%%%%%%%%%%%%%%%%%%%%%%%%%%%%%%%%%%%%%%%%%%%%%%%%%%%%%%%%%%%%%%%%%%%%%%%
%%%%%
%%%%           Graphs for Delta A_CP
%%%%%
%%%%%%%%%%%%%%%%%%%%%%%%%%%%%%%%%%%%%%%%%%%%%%%%%%%%%%%%%%%%%%%%%%%%%%
\FIGURE{\epsfig{file=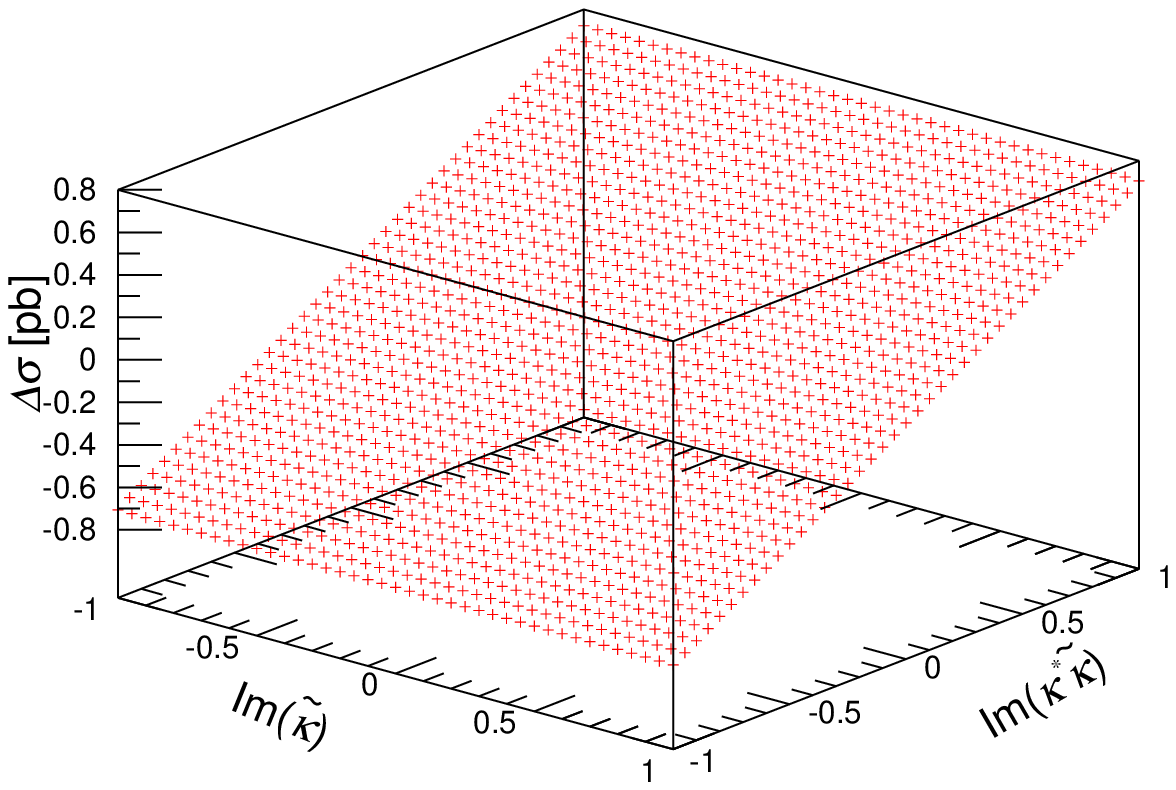,width=10cm}
\caption{Surface plots displaying the dependence of
$\Delta \sigma$ on Im($\kappa^* \ktil$) and Im($\ktil$) 
at$\sqrt{s}$=2.0 TeV.}
\label{fig:cp1}}
%%%%%%%%%%%%%%%%%%%%%%%%%%%%%%%%%%%%%%%%%%%%%%%%%%%%%%%%%%%%%%%%%%%%%

%%%%%%%%%%%%%%%%%%%%%%%%%%%%%%%%%%%%%%%%%%%%%%%%%%%%%%%%%%%%%%%%%%%%%%%
%%%
%%%              Table of dependence of polarization asymmetry
%%%
%%%%%%%%%%%%%%%%%%%%%%%%%%%%%%%%%%%%%%%%%%%%%%%%%%%%%%%%%%%%%%%%%%%%%%%
\begin{table}
\begin{center}
\hspace*{-0.7cm}
{\scriptsize
\begin{tabular}{|c|c||c|c|c|c|c|c|c|}
\hline
 \multicolumn{2}{|c|}{$\Delta \sigma$}&\multicolumn{7}{c|}{Im($\ktil$)}\\
\cline{3-9}
\multicolumn{2}{|r|}{\tiny[Pb]} 
                        &1.0&0.1&0.01&0& $-$0.01& $-$0.1  & $-$1.0  \\ 
\hline \hline
            &1.0  &7.06$\times 10^{-1}$&7.13$\times 10^{-1}$
                &7.14$\times 10^{-1}$&7.14$\times 10^{-1}$
                &7.14$\times 10^{-1}$&7.15$\times 10^{-1}$
                &7.23$\times 10^{-1}$  \\                         \cline{2-9} 
          &0.5  &3.48$\times 10^{-1}$&3.56$\times 10^{-1}$
                &3.57$\times 10^{-1}$&3.57$\times 10^{-1}$
                &3.57$\times 10^{-1}$&3.58$\times 10^{-1}$
                &3.67$\times 10^{-1}$  \\                          \cline{2-9}
           &0.1&6.28$\times 10^{-2}$&7.06$\times 10^{-2}$
                &7.13$\times 10^{-2}$&7.14$\times 10^{-2}$
                &7.15$\times 10^{-2}$&7.23$\times 10^{-2}$
                &8.00$\times 10^{-2}$  \\                           \cline{2-9}
%          &0.05&2.71$\times 10^{-2}$&3.48$\times 10^{-2}$
%                &3.56$\times 10^{-2}$&3.57$\times 10^{-2}$
%                &3.58$\times 10^{-2}$&3.66$\times 10^{-2}$
%                &4.43$\times 10^{-2}$  \\                          \cline{2-9}
           &0.01&$-$1.49$\times 10^{-3}$&6.28$\times 10^{-3}$
                &7.06$\times 10^{-3}$&7.14$\times 10^{-3}$
                &7.23$\times 10^{-3}$&8.00$\times 10^{-3}$
                &1.58$\times 10^{-2}$  \\                           \cline{2-9}
 Im($\kappa^*\ktil$)
           &0&$-$8.62$\times 10^{-3}$&$-$8.62$\times 10^{-4}$
                &8.62$\times 10^{-5}$&              0
                &8.62$\times 10^{-5}$&8.62$\times 10^{-4}$
                &8.62$\times 10^{-3}$  \\                           \cline{2-9}
          &$-$0.01&$-$1.58$\times 10^{-2}$&$-$8.00$\times 10^{-3}$
                &$-$7.23$\times 10^{-3}$&$-$7.14$\times 10^{-3}$
                &$-$7.06$\times 10^{-3}$&6.28$\times 10^{-3}$
                &1.49$\times 10^{-3}$  \\                           \cline{2-9}
%          &$-$0.05&$-$4.43$\times 10^{-2}$&$-$3.66$\times 10^{-2}$
%                &$-$3.58$\times 10^{-2}$&$-$3.57$\times 10^{-2}$
%                &$-$3.56$\times 10^{-2}$&$-$3.48$\times 10^{-2}$
%                &$-$2.71$\times 10^{-2}$  \\                       \cline{2-9}
          &$-$0.1&$-$8.00$\times 10^{-2}$&$-$7.23$\times 10^{-2}$
                &$-$7.15$\times 10^{-2}$&$-$7.14$\times 10^{-2}$
                &$-$7.13$\times 10^{-2}$&$-$7.06$\times 10^{-2}$
                &6.28$\times 10^{-2}$  \\                           \cline{2-9}
          &$-$0.5&$-$3.67$\times 10^{-1}$&$-$3.58$\times 10^{-1}$
                &$-$3.57$\times 10^{-1}$&$-$3.57$\times 10^{-1}$             
                &$-$3.57$\times 10^{-1}$&$-$3.56$\times 10^{-1}$
                &$-$3.48$\times 10^{-1}$  \\                       \cline{2-9}
          &$-$1.0&$-$7.23$\times 10^{-1}$&$-$7.15$\times 10^{-1}$
                &$-$7.14$\times 10^{-1}$&$-$7.14$\times 10^{-1}$
                &$-$7.14$\times 10^{-1}$&$-$7.13$\times 10^{-1}$
                &$-$7.06$\times 10^{-1}$  \\                      \cline{2-9}

\hline
\end{tabular}
}
\end{center}
\caption{The dependence of the polarization symmetry 
$\Delta \sigma$ [pb] on Im($\kappa^* \ktil$) and Im($\ktil$).}
\label{tab:cpa}
\end{table}
%%%%%%%%%%%%%%%%%%%%%%%%%%%%%%%%%%%%%%%%%%%%%%%%%%%%%%%%%%%%%%%%%%%%%%%%%%%%

We show the dependence of $\Delta \sigma$ on 
Im($\kappa^* \ktil$) and Im($\ktil$) at the upgraded Tevatron Energy 
($\sqrt{s}$=2.0 TeV) in  Fig.~{\ref{fig:cp1}}.
In addition, the magnitude of  $\Delta \sigma$ for
some combinations of the magnitude of Im($\kappa^* \ktil$) and Im($\ktil$)
are listed in Table~\ref{tab:cpa}.
In our numerical calculation, we used the parton distribution functions of
CTEQ5M~\cite{cteq} and as input data $\mt$=174.3 GeV~\cite{pdg}. 
In  Fig.~\ref{fig:cp1}, 
we can easily see that 
$\Delta \sigma$  strongly depends on Im($\kappa^* \ktil$)
though we cannot individually  determine these values.
This dependence can be also seen in the same approach to  
$C\!P$-violation for 
$e^+ e^- \to t \bar{t} {\rm g}$~\cite{rindani}.

Secondly, let us estimate the $C\!P$-violating
observable ${\cal A}_{cp}$  which is  defined by 
Eq.~\eqref{eq:crossasym};
${\cal A}_{cp} \equiv {\Delta \sigma}/{\sigma_{\rm total}}$.
Since the total cross section $\sigma_{\rm total}$  calculated
 using~\eqref{eq:efflag} has many  undetermined 
parameters such as $|\kappa|^2$, $|\ktil|^2$ and  ${\rm Re}({\kappa})$,
we approximated  $\sigma_{\rm total}$ by
$\sigma_{SM}$ predicted by the SM.
This approximation is not unreasonable because the cross section
given by the SM is in good agreement with experimental result.
In order to observe ${\cal A}_{cp}$ at the 90\% Confidence Level (CL),  
${\cal A}_{cp}$ must satisfy the condition:
\begin{equation}
\label{eq:erra}
|{\cal A}_{cp}|\ge \frac{1.64}{\sqrt{N_{\rm events}}}, 
\end{equation}
where $N_{\rm events}$ means the total number of events which can be
experimentally reconstructed.

At the upgraded Tevatron with center-of-mass energy $\sqrt{s}$ = 2.0
TeV, the decay mode which is  most sensitive to
the helicities of the produced top and anti-top quark,
is the dilepton mode,
$t \bar{t} \rightarrow b \bar{b} W^+(\to l^+ \nu_l) W^-(\to l^- \bar{\nu}_l),$
because the energies of leptons produced from $W$ bosons 
and $b$ ($\bar{b}$) are sensitive to the helicities of the produced 
top and anti-top quark as mentioned in Section~\ref{sec:cpo}.  
The total number of events of the dilepton mode are expected to be
detected about 1200 events~\cite{topevent}~\footnote{
In this work, we focus only on Run I\!I\!I (TeV 33) experiment
due to the statistical advantage,
though 
there is also Run I\!I experiment.}.
Therefore, Eq.~\eqref{eq:erra} becomes
\begin{align}
\label{eq:erra2}
|{\cal A}_{cp}|&=\left|\frac{\Delta \sigma}{\sigma_{{}_{SM}}}\right|
\ge \frac{1.64}{\sqrt{N_{\rm events}}}=\frac{1.64}{\sqrt{1200}}.
\end{align}
Then, with $\sigma_{SM}=7.5$pb~\cite{topevent},
the $\Delta \sigma$ should satisfy the relation:
\begin{equation}
 \left|\Delta \sigma \right|  \ge 0.35 [{\rm pb}].
\end{equation}
%%%%%%%%%%%%%%%%%%%%%%%%%%%%%%%%%%%%%%%%%%%%%%%%%%%%%%%%%%%%%%%%%%%%%%%
%%%
%%%              Table of dependence of polarization asymmetry
%%%
%%%%%%%%%%%%%%%%%%%%%%%%%%%%%%%%%%%%%%%%%%%%%%%%%%%%%%%%%%%%%%%%%%%%%%%
\begin{table}
\begin{center}
\hspace*{-0.7cm}
{\small
\begin{tabular}{|c|c||c|c|c|c|c|c|c|}
\hline
 \multicolumn{2}{|c|}{$\Delta {\cal A}_{cp}$}
&\multicolumn{7}{c|}{Im($\ktil$)}\\
\cline{3-9}
\multicolumn{2}{|r|}{[$\times 10^{-2}$]} 
                        &1.0&0.1&0.01&0& $-$0.01& $-$0.1  & $-$1.0  \\ 
\hline \hline
          &1.0  &$9.41$&$9.51$
                &$9.52$&$9.52$
                &$9.52$ &$9.52$
                & $9.63$\\                         \cline{2-9} 
          &0.7  &$6.55$&$6.65$
                &$6.66$&$6.67$
                &$6.67$ &$6.68$
                & $6.78$\\                         \cline{2-9} 
          &0.5  &$\times$&$4.75$
                &$4.76$&$4.76$
                &$4.76$ &$4.77$
                & $4.88$\\                         \cline{2-9}
          &0.1&$\times$&$\times$
                &$\times$&$\times$
                &$\times$&$\times$
                &$\times$  \\                          \cline{2-9}
 Im($\kappa^*\ktil$)
              &0&$\times$&$\times$
                &$\times$&$\times$
                &$\times$&$\times$
                &$\times$ \\                           \cline{2-9}
         &$-$0.1&$\times$&$\times$
                &$\times$&$\times$
                &$\times$&$\times$
                &$\times$         \\                       \cline{2-9}
         &$-$0.5&$-4.68$&$-4.77$
                &$-4.76$&$-4.76$
                &$-4.76$&$-4.75$
                &$\times$\\                           \cline{2-9}
         &$-$0.7&$-6.78$&$-6.68$
                &$-6.67$&$-6.67$
                &$-6.66$&$-6.65$
                &$-6.55$\\                           \cline{2-9}
         &$-$1.0&$-9.63$&$-9.52$
                &$-9.52$&$-9.52$
                &$-9.52$&$-9.51$
                &$-9.41$\\                           \cline{2-9}

\hline
\end{tabular}
}
\end{center}
\caption{
The expected magnitude of the ${\cal A}_{cp}$ at the Upgraded Tevatron.
A symbol ``$\times$'' means that the magnitude of the  ${\cal A}_{cp}$
are smaller than that at 90\% CL as is defined by Eq.(\ref{eq:erra}).
} 
\label{tab:acp}
\end{table}
%%%%%%%%%%%%%%%%%%%%%%%%%%%%%%%%%%%%%%%%%%%%%%%%%%%%%%%%%%%%%%%%%%%%%%%%%%%%
In Table~\ref{tab:acp},
we present the magnitude of ${\cal A}_{cp}$
at 90 \% CL
for the typical combinations of the magnitude
of Im($\kappa^* \ktil$) and Im($\ktil$).
It is interesting that we can observe 
the  ${\cal A}_{cp}$  at 90 \% CL 
when $|$Im($\kappa^* \ktil$)$|>$0.5,
even if Im($\ktil$)=0, as shown in Table~\ref{tab:acp}.

Even if we do not observe signals of ${\cal A}_{cp}$ 
at the upgraded Tevatron,
we can obtain the constraint of the Im($\kappa^* \ktil$) and
Im($\ktil$).

\section{SUMMARY AND OUTLOOK}
\label{sec:summary}
Expecting the observation of  $p\bar{p}\to t \bar{t} X$ process at
the upgraded Tevatron,
we estimated the magnitude of 
$C\!P$-violation  effect by using an effective Lagrangian
which includes the anomalous chromomagnetic
and chromoelectric coupling. 
Our analysis does not depend on the specific  models because we took 
the anomalous chromomagnetic and chromoelectric
coupling as a free parameter. 
We found that $C\!P$-violation depends only on
Im($\kappa^* \ktil$) and Im($\ktil$).
Especially, 
dependence of  Im($\kappa^* \ktil$)  is stronger than 
that of Im($\ktil$) in this process. 
Furthermore, we pointed out that  $C\!P$-violating 
observables, ${\cal A}_{cp}$, 
can be measured at 90 \% CL in some combinations of 
the magnitude of Im($\kappa^* \ktil$) and Im($\ktil$).

Though the same analysis can be applied for  the Large Hadron collider (LHC)
at CERN which will start in 2005, 
the behavior of the  $C\!P$-violation may be different from 
this analysis 
because in this case
the dominant process of top quark pair production is
two-gluon fusion which was neglected in this work.

Since this analysis was  done without  specific models,
the model could not be specified from this analysis,
even if the $C\!P$-violation is  measured in this  process.
However, such  measurements could  give the constraints
on  parameters in any models.

Finally, we did  not analyze decays 
of top quarks in process, which needs further investigation.
The analysis of $C\!P$-violation at the LHC is
also important and   will be done further in future work.

\section*{Acknowledgements}
I wish to express my gratitude to T. Morii for careful reading of 
this manuscript and valuable comments. 
I am also grateful to Z. Hioki, C. S. Lim and S. Oyama for
many useful discussion related to this work. 
Finally, I am thankful to S. Kim  for sharing information about 
the upgraded Tevatron.
\appendix
\section{THE SPIN VECTORS AND FOUR--MOMENTA IN PARTON CENTER-OF-MASS SYSTEM} 
\label{sec:cms}
The definition of the spin vectors and four-momenta in 
the parton center-of-mass system is given in this appendix.
\\
\\
We define the spin four-vectors, $\st,\stb$, in the parton center-of-mass
system (CMS) in terms of a spin angle $\xi$
as it is illustrated in Ref~\cite{system1}.
\begin{align}
&\st=\frac{1}{\gamma_t}(\beta_t\cos\xi;\gamma_t\sin\xi,0,-\cos\xi)\\
&\stb=\frac{1}{\gamma_t}(\beta_t\cos\xi;\gamma_t\sin\xi,0,\cos\xi)
\end{align}
with $\beta_t \equiv \sqrt{1-4m_t^2/\hat{s}}$, 
$\gamma_t\equiv\sqrt{1-\beta_t^2}$.

Here, we set as $\xi$=$\pi$ because
we calculated in the helicity basis.

In the parton CMS with the z-axis chosen to be along the top quark
direction of motion, the four-momenta read as follows:
\begin{align}
&\pt=\frac{\sqrt{\shat}}{2}(1;0,0,\beta_t) \non \\
&\ptb=\frac{\sqrt{\shat}}{2}(1;0,0,-\beta_t) \non \\
&\pa=\frac{\sqrt{\shat}}{2}(1;\sin\htheta,0,\cos\htheta)\non \\
&\pb=\frac{\sqrt{\shat}}{2}(1;-\sin\htheta,0,-\cos\htheta),
\end{align}
where $\hat{\theta}$ denotes the emission angle of the top quark.

%%%%%%%%%%%  BIBLIOGRAPHY  %%%%%%%%%%%%

\end{document}